\documentclass[%
showkeys,
 amsmath,amssymb,
]{revtex4-1}

\usepackage{graphicx}
\usepackage{dcolumn}
\usepackage{bm}

\usepackage{CJK}

\usepackage{color}
\usepackage{amssymb}
\usepackage{amsfonts}

\usepackage[colorlinks,urlcolor=blue,linkcolor=blue,citecolor=blue,anchorcolor=blue]{hyperref}

\begin{document}

\preprint{APS/123-QED}

\title{Automatic Fourier transform and self-Fourier beams due to parabolic potential}

\author{Yiqi Zhang$^1$}
\email{zhangyiqi@mail.xjtu.edu.cn}
\author{Xing Liu$^1$}
\author{Milivoj R. Beli\'c$^{2}$}
\author{Weiping Zhong$^3$}
\author{Milan S. Petrovi\'c$^4$}
\author{Yanpeng Zhang$^{1}$}
\email{ypzhang@mail.xjtu.edu.cn}
\affiliation{%
 $^1$Key Laboratory for Physical Electronics and Devices of the Ministry of Education \& Shaanxi Key Lab of Information Photonic Technique,
Xi'an Jiaotong University, Xi'an 710049, China \\
$^2$Science Program, Texas A\&M University at Qatar, P.O. Box 23874 Doha, Qatar \\
$^3$Department of Electronic and Information Engineering, Shunde Polytechnic, Shunde 528300, China \\
$^4$Institute of Physics, P.O. Box 68, 11001 Belgrade, Serbia
}%

\date{\today}

\begin{abstract}
  \noindent
We investigate the propagation of light beams
  including Hermite-Gauss, Bessel-Gauss and finite energy Airy beams
  in a linear medium with parabolic potential.
  Expectedly, the beams undergo oscillation during propagation,
  but quite unexpectedly they also perform automatic Fourier transform,
  that is, periodic change from the beam to its Fourier transform and back.
  The oscillating period of parity-asymmetric beams is twice that of the parity-symmetric beams.
  In addition to oscillation, the finite-energy Airy beams exhibit periodic inversion during propagation.
  Based on the propagation in parabolic potential, we introduce a class of optically-interesting beams that
  are self-Fourier beams -- that is, the beams whose Fourier transforms are the beams themselves.
\end{abstract}

\keywords{parabolic potential, self-Fourier beam}
\maketitle

%

\section{Introduction}
It is well know that a light beam undergoes discrete diffraction while propagating in a waveguide array,
and that such a diffraction is prohibited when the refractive index of the waveguide array is appropriately modulated \cite{lederer.pr.463.1.2008,garanovich.pr.518.1.2012}.
The phenomenon is linear, that is, obtained without invoking nonlinearity in the paraxial wave equation \cite{peschel.ol.23.1701.1998}.
Likewise, in free space or a linear bulk medium a light beam will diffract unless it belongs to the family of nondiffracting beams
\cite{siviloglou.ol.32.979.2007, kaminer.prl.108.163901.2012} -- a class of linear beams that attracted a lot of attention in the past few years.
A celebrated member of this class is the Airy beam \cite{siviloglou.ol.32.979.2007, siviloglou.prl.99.213901.2007,hang.pra.90.023822.2014,zhang.rrp.67.1099.2015,diebel.oe.23.24351.2015,zhong.oe.23.23867.2015}.
In optics, ways to effectively modulate light beams have always been
high on  research agenda. As an effective tool, a  photonic
potential -- a ``potential'' embedded in the medium's index of
refraction -- is often used in linear optics and extensively
referenced in the literature. It comes in different forms, as
exemplified by vastly different photonic crystal structures.
A linear potential which affects the properties of an Airy plasmon
beam and is used to control acceleration of Airy beams has been
reported in \cite{liu.ol.36.1164.2011,ye.ol.36.3230.2011}. An
external longitudinally-dependent transverse potential will modulate
the propagating trajectory of the light beam according to the form
of the potential \cite{efremidis.ol.36.3006.2011}. In Refs.
\cite{bandres.oe.15.16719.2007,zhang.oe.23.10467.2015,zhang.ol.40.3786.2015} authors
discussed the propagation and transformation of (finite-energy) Airy
beams in a linear medium with a parabolic potential, in which
periodic inversion, oscillation, and phase transition were reported.
One should recall that in a parabolic potential,
light propagation is equivalent to a fractional Fourier transform (FT)
\cite{zhou.apb.109.549.2012}, the fact first noted by
Mendlovic and his collaborators
\cite{mendlovic.josaa.10.1875.1993,mendlovic.ao.33.6188.1994}. The
fractional FT can be conveniently introduced through the propagation
in gradient-index (GRIN) media, which can further be connected with
the propagation in the parabolic potential and the automatic FT --
as done in this article. By now, beam propagation in GRIN media
has matured to a proper part of optics that is best surveyed from a
dedicated monograph \cite{gomez.book.2002}. In a nonlinear medium,
the strongly nonlocal nonlinearity can be cast into a parabolic-like
potential \cite{snyder.science.276.1538.1997} and the corresponding
modulation effects more easily investigated
\cite{lu.pra.78.043815.2008,zhou.lpl.11.105001.2014,shen.sr.5.9814.2015}.

In this article, we investigate the light beam management by a
parabolic potential in a linear medium, theoretically and
numerically. The program presented extends the traditional Fourier
optics, which deals with the free-space propagation, to the realm of
propagation in a parabolic potential. Since such a potential causes
harmonic oscillation of light beams, it is interesting to
investigate the influence of the potential on the dynamics of useful
light beams, such as Hermite-Gauss (HG), Bessel-Gauss (BG), and
finite energy Airy beams (ABs). Even though there are many papers
dealing with the parabolic potential and linear harmonic
oscillation, we believe that results obtained here have not been
reported before, to the best of our knowledge.

The organization of the article is as follows. In Sec. \ref{model},
we describe the problem and introduce the theoretical model of beam
propagation. In Sec. \ref{discussion}, we discuss the repercussions
of the model on the dynamics of the beams mentioned. In Sec.
\ref{ana_solu}, we solve the model, obtain analytical solutions, and
analyze those solutions. Based on the theoretical model, we discover
a class of interesting self-Fourier beams and present them in Sec.
\ref{fourier}.
Section \ref{conclusion} concludes the paper.

\section{Theoretical model}\label{model}

The paraxial propagation of a beam in a linear medium with an external parabolic potential,
is described by the dimensionless Schr\"odinger equation
\begin{equation}\label{eq1}
   i\frac{\partial \psi}{\partial z} +\frac{1}{2} \frac{\partial^2 \psi}{\partial x^2} -\frac{1}{2} \alpha^2 x^2 \psi=0,
\end{equation}
where $x$ and $z$ are the normalized transverse coordinate and the
propagation distance, respectively, scaled by some transverse width
$x_0$ and the corresponding Rayleigh range $kx_0^2$. Here, $k=2\pi
n/\lambda_0$ is the wavenumber, $n$ the index of refraction, and
$\lambda_0$ the wavelength in vacuum. Parameter $\alpha$ scales the
width of the potential. For our purposes, the values of parameters
can be taken as $x_0=100\,\mu \rm m$, $n=1.45$, and
$\lambda_0=600\,\rm nm$ \cite{zhang.ol.38.4585.2013,
zhang.oe.22.7160.2014}.

Equation (\ref{eq1}) has many well-known solutions; we will utilize
the ones that are of interest in the paraxial beam propagation. But
before selecting any solutions, we perform the Fourier transform
(FT) of Eq. (\ref{eq1}), to obtain
\begin{equation}\label{eq2}
   i\frac{\partial \hat{\psi}}{\partial z} + \frac{1}{2}\alpha^2 \frac{\partial^2 \hat{\psi}}{\partial k^2} - \frac{1}{2} k^2 \hat{\psi}=0,
\end{equation}
where the FT is defined as
\[
\hat{\psi} = \int^{+\infty}_{-\infty} \psi e^{-i k x} dx, \quad
\psi = \frac{1}{2\pi}\int^{+\infty}_{-\infty} \hat{\psi} e^{i k x} dk.
\]
Obviously, Eq. (\ref{eq2}) can be put in the same form as Eq.
(\ref{eq1}), especially if $\alpha=1$. Both equations can have the
same solutions but expressed in different spaces, real and inverse.
The eventual differences in solutions come to the fore once the
equations are presented as boundary-value problems in specific
physical settings. Also, Eqs. (\ref{eq1}) and (\ref{eq2}) indicate
that localized light beams in real ($x$) and inverse ($k$) spaces
may share the same dynamics from a mathematical point of view.
Therefore, it is natural to consider the following scenario in
linear optical setting for a beam propagating according to Eq.
(\ref{eq1}): (i) The beam undergoes FT during propagation, and then
experiences an inverse FT, to reconstruct the initial beam; (ii) The
process represented in (i) occurs periodically during propagation,
and profile oscillation and recurrence take place along the
propagation direction.

\begin{figure}[htbp]
\centering
  \includegraphics[width=0.5\columnwidth]{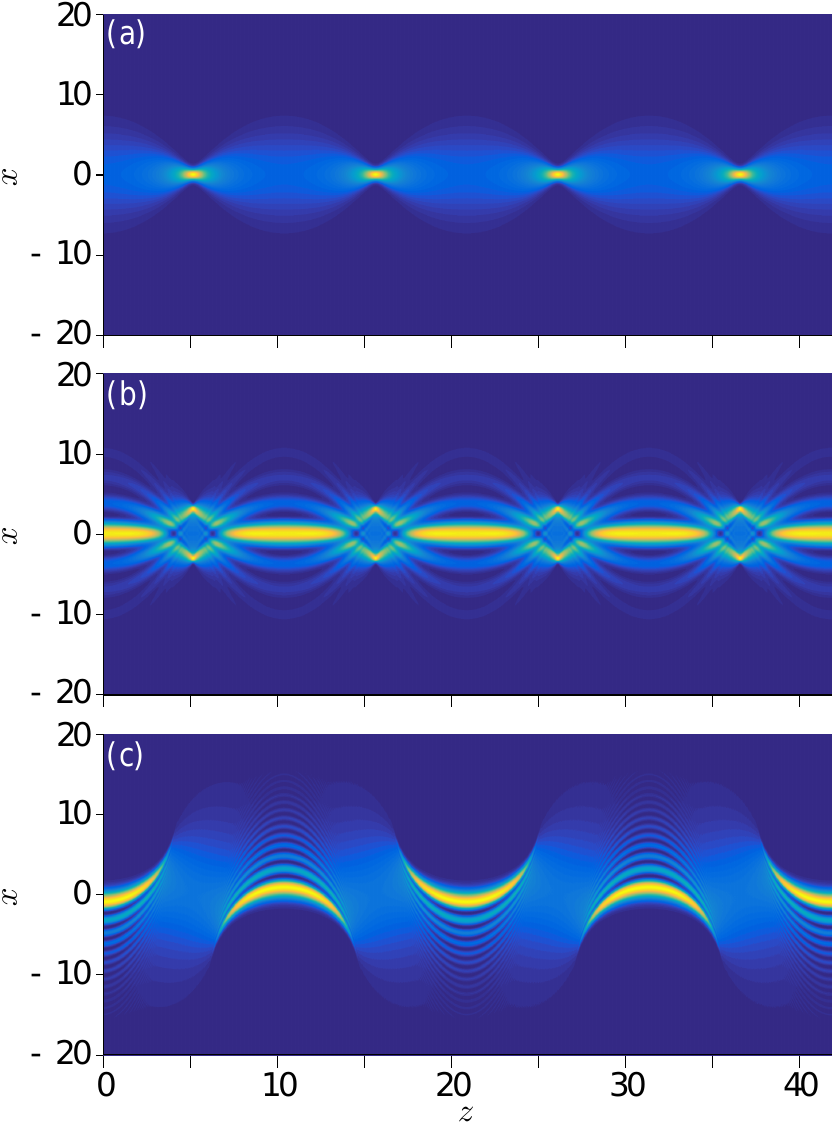}
  \caption{(Color online) Propagation of beams in a linear medium with parabolic potential.
  (a) Hermite-Gauss beam, with input $\psi(x)=\exp(-x^2/2\sigma^2)H_0(x/\sigma)$ with $\sigma=5$.
  (b) Bessel-Gauss beam, input $\psi(x)= \exp(-x^2/2\sigma^2)J_0(x)$ with $\sigma=10$.
  (c) Finite energy Airy beam, input $\psi(x) = \exp(ax){\rm Ai}(x)$.
  The parameters are $a=0.1$ and $\alpha=0.3$.
  }
  \label{fig1}
\end{figure}

The scenario outlined resembles a repeating $4f$ correlator system
from Fourier optics, except that it deals with the propagation in a
parabolic potential.
To present a realization of the scenario outlind, we display the
propagation of HG, BG and a finite energy AB in Figs.
\ref{fig1}(a)-\ref{fig1}(c), respectively. Clearly, the parabolic
potential imposes simple harmonic oscillation on an HG beam, as
shown in Fig. \ref{fig1}(a), prevents discrete-like diffraction of a
BG beam (Fig. \ref{fig1}(b)), and causes periodic inversion of an AB
(Fig. \ref{fig1}(c)). This inversion
\cite{bandres.oe.15.16719.2007,zhang.oe.23.10467.2015} is different
from the inversion of an AB reported in fibers, when the third-order
dispersion is taken into account \cite{driben.ol.38.2499.2013}, and
from the double focusing of an AB in Ref.
\cite{rogel-salazar.pra.89.023807.2014}.

\begin{figure*}[htbp]
  \centering
  \includegraphics[width=\textwidth]{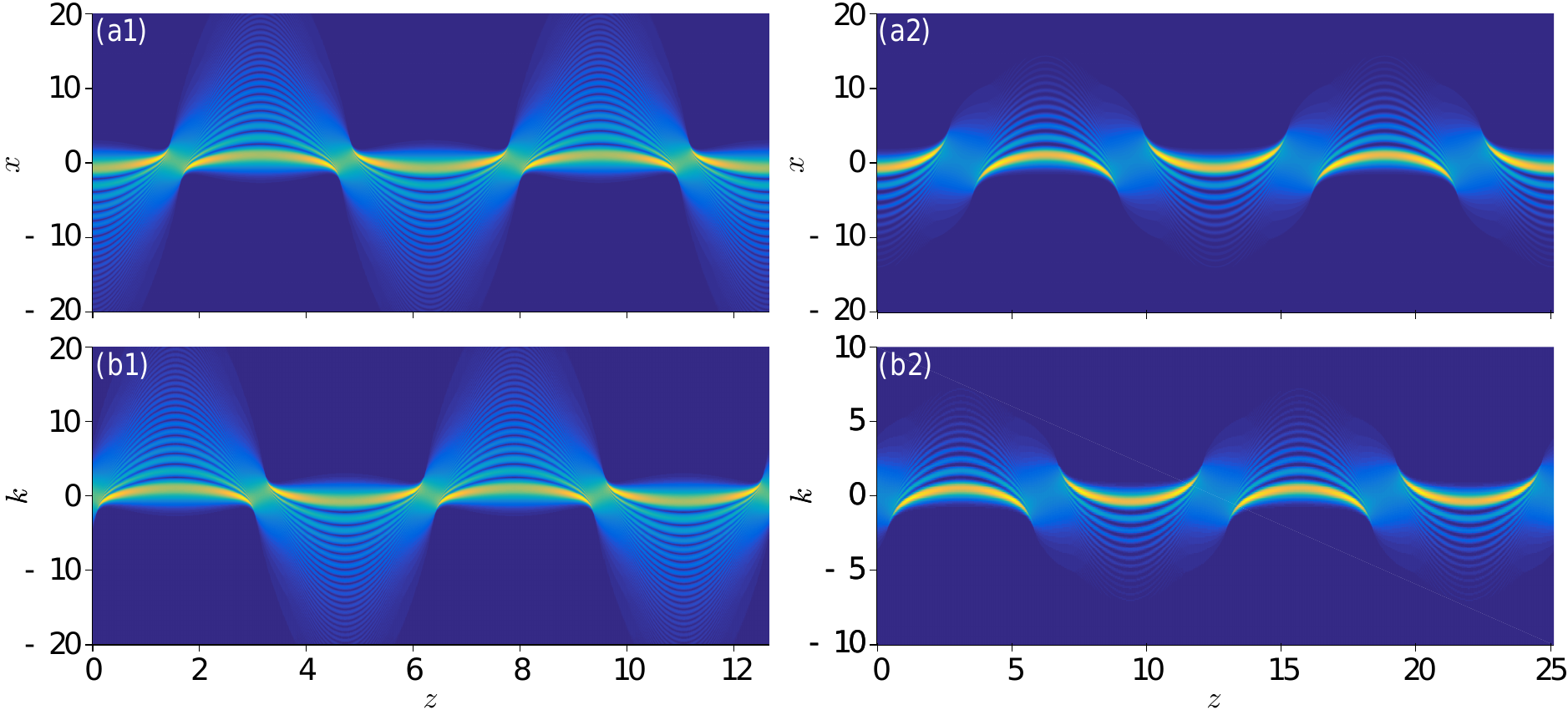}
  \caption{(Color online) Periodic inversion and automatic Fourier transform of a finite energy Airy beam
  during propagation.
  (a) Real space.
  (b) Inverse space.
  The parameters are: $a=0.1$, $\alpha=1$ and $\alpha=0.5$ for the left panels and the right panels, respectively.
  }
  \label{fig2}
\end{figure*}

In Figs. \ref{fig1}(a) and \ref{fig1}(b), the initial beams are
parity-symmetric, so an odd-integer multiple of half periods
$\mathcal{D}_s$ of harmonic oscillation is needed for the beam to
realize its FT. However, in Fig. \ref{fig1}(c) the initial beam is
parity-asymmetric, so it inverts once before reconstruction; as a
result, the relation between oscillation periods of symmetric and
asymmetric beams is  $\mathcal{D}_{\rm as}=2\mathcal{D}_s$. The
corresponding FTs are located at an integer multiple of $(2m-1)
\mathcal{D}_{\rm as}/4$, with $m$ being an even integer. If $m$ is
an odd integer, the rule gives the FT of the finite energy AB at
inversion. This demonstrates that HG, BG, and finite energy ABs
perform an automatic Fourier transformation, which means that the FT
of these beams will appear automatically and periodically during
propagation, based on this simple model.

\section{Discussion}\label{discussion}

In this section, we address the properties of the three beams during
propagation. Specifically, in Fig. \ref{fig2} we depict the periodic
inversion and automatic FT of an AB during propagation. In Figs.
\ref{fig2}(a1) and \ref{fig2}(b1), $\alpha=1$, so that the beam
intensities in the real domain and the inverse domain are identical.
In Figs.  \ref{fig2}(a2) and \ref{fig2}(b2), $\alpha=0.5$;
therefore, there is a transverse scaling between the intensities in
real and inverse domains. Using Parseval's theorem
\[
\int_{-\infty}^{+\infty} |\psi(x)|^2dx = \frac{1}{2\pi} \int_{-\infty}^{+\infty} |\hat{\psi}(k)|^2 dk
\]
and a gauge transformation, we can obtain the FT of the three initial beams in real space.

\begin{figure}[htbp]
  \centering
  \includegraphics[width=0.5\columnwidth]{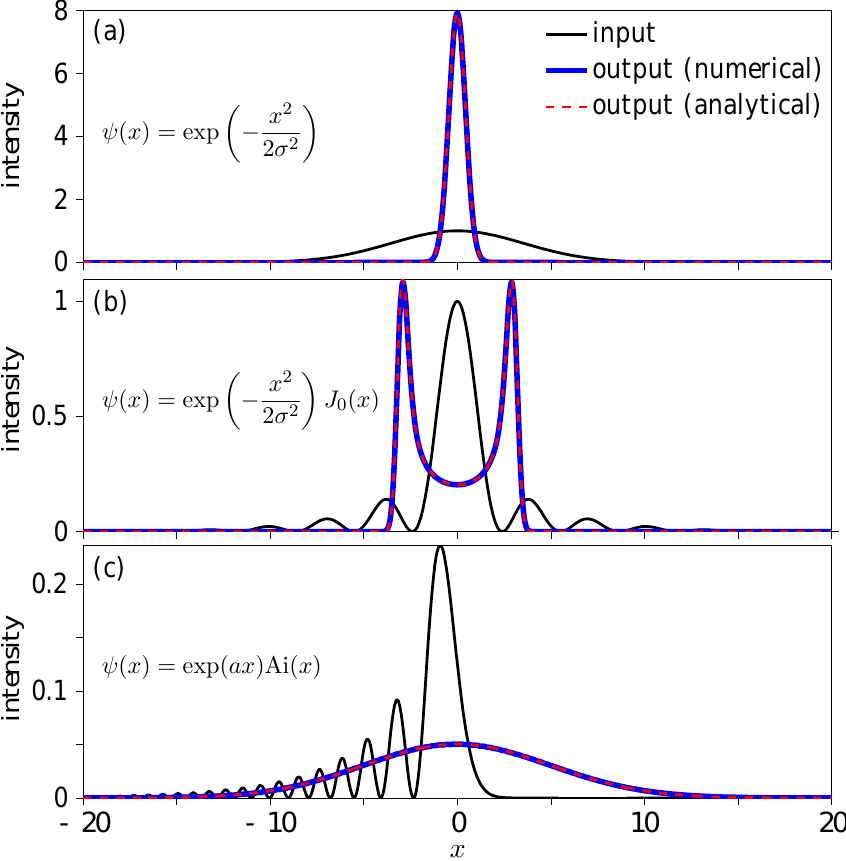}
  \caption{(Color online) Comparison between numerical and analytical results.
  (a)-(c) Corresponding to Figs. \ref{fig1}(a)-\ref{fig1}(c), respectively;
  outputs are taken at $z=\mathcal{D}_s/2$, $\mathcal{D}_s/2$, and $3\mathcal{D}_{\rm as}/4$.
  Perfect agreement is visible.}
  \label{fig3}
\end{figure}

For an arbitrary order HG beam $\psi(x,\,0)=\exp(-x^2/2\sigma^2)H_n(x/\sigma)$,
with $n$ the order of Hermite polynomials [in Fig. \ref{fig1}(a), the order is 0],
\begin{subequations}
\begin{align}\label{eq3a}
  \psi\left(x,\,z = \frac{m}{2}\mathcal{D}_s \right) =
   (-i)^n \sqrt{\alpha} \sigma
   \exp\left( -\frac{1}{2}\alpha^2 \sigma^2 x^2 \right) H_n \left( \alpha\sigma x \right),
\end{align}
where $m$ is an odd integer. If we let $\alpha=1$ and $\sigma=1$, we
find Eq. (\ref{eq3a}) to coincide completely with the initial beam.
The reason is that $\psi(x,\,0)=\exp(-x^2/2)H_n(x)$ is an eigenmode
of Eq. (\ref{eq1}), which is a well known fact.

Concerning the BG beam, we also consider the general case
$\psi(x,\,0)=\exp(-x^2/2\sigma^2)J_n(x)$, with $n$ being the order of the Bessel function.
The corresponding FT can be obtained as a convolution
of FTs of the Gauss and Bessel functions.
After some algebra, one obtains
\begin{align}\label{eq3b}
  \psi \left(x,\,z=\frac{m}{2}\mathcal{D}_s \right) = (-i)^n \frac{\sigma \sqrt{\alpha}}{\pi}
  \int^{\pi/2}_{-\pi/2} T_n(\sin\theta) \exp\left[-\frac{\sigma^2}{2} \left(\alpha x - \sin \theta\right)^2 \right] d\theta,
\end{align}
where $T_n$ is the Chebyshev polynomial of the first kind.
For the case in Fig. \ref{fig1}(b), $n=0$.

The case of an AB is more interesting, but also more involved.
Since the FT of
$\psi(x) = {\rm Ai}(x)\exp(ax)$ is
\[{\hat \psi} (k) = \exp(-ak^2)\exp\left[ \frac{a^3}{3} + \frac{i}{3}\left(k^3-3a^2k\right)\right],\]
the beam envelope can be written as
\begin{align}\label{eq3c}
\psi \left( x,\,z=  \frac{2m-1}{4}\mathcal{D}_{\rm as} \right)
   & = \sqrt{-\frac{s\alpha}{2\pi}} \exp\left(-\alpha^2 a x^2 + i\frac{\pi}{4} \right) 
    \exp \left[ \frac{a^3}{3} + i \frac{s}{3} \left( \alpha^3 x^3 - 3\alpha a^2x \right) \right].
\end{align}
In Eq. (\ref{eq3c}), $s=1$ if $m$ is odd and $s=-1$ if $m$ is even.
\end{subequations}
It will be verified later in the article that the oscillation period is
\begin{subequations}
\begin{equation}\label{eq4a}
\mathcal{D}_s=\frac{\pi}{\alpha}
\end{equation}
for parity symmetric beams (e.g., HG and BG), and
\begin{equation}\label{eq4b}
\mathcal{D}_{\rm as}=\frac{2\pi}{\alpha}
\end{equation}
for beams lacking parity symmetry (e.g., ABs).
\end{subequations}

The analytical solutions given by Eqs. (\ref{eq3a})-(\ref{eq3c})
completely agree with the numerical simulations displayed in Figs.
\ref{fig1} and \ref{fig2}, as shown in Fig. \ref{fig3}. From Eqs.
(\ref{eq3a})-(\ref{eq3c}), (\ref{eq4a}) and (\ref{eq4b}), one can
see that the harmonic oscillation, self-Fourier transform and
periodic inversion are only dependent on $\alpha$; apodization
factors $\sigma$ and $a$ do not affect these phenomena.

Thus far, we have found analytical expressions of propagating beams
at certain distances; the procedure may also be applied to other
kinds of initial light beams, regardless of their parity symmetry.
The question is, what happens in between these specific distances.

\section{Analytical solutions}\label{ana_solu}

We now go back to Eq. (\ref{eq1}); as mentioned, it possesses many
solutions. We select ones that are relevant for the study at hand,
some of which are derived by the self-similar method
\cite{ponomarenko.prl.97.013901.2006, zhong.pra.75.061801.2007,
yang.ctp.53.937.2010}. Generally, the solution of Eq. (\ref{eq1})
can be written as \cite{bernardini.epj.16.58.1995,
kovalev.josaa.31.914.2014,
bandres.oe.15.16719.2007,zhang.oe.23.10467.2015}
\begin{equation}\label{eq5}
  \psi(x,z) = \int_{-\infty}^{+\infty} \psi(\xi,0) \sqrt{\mathcal{H}(x,\xi,z)} d\xi,
\end{equation}
where
\begin{align}\label{eq6}
  \mathcal{H}(x,\xi,z) =  -\frac{i}{2\pi}\alpha\csc\left(\alpha z\right)
         \exp \left\{ i\alpha \cot\left(\alpha z\right)  \left[ x^2+\xi^2-2x\xi\sec\left(\alpha z\right) \right] \right\}.
\end{align}
Combining Eqs. (\ref{eq5}) and (\ref{eq6}), after some algebra one ends up with
\begin{align}\label{eq7}
  \psi(x,z) = f(x,z)
  \int^{+\infty}_{-\infty} \left[\psi(\xi,0)\exp\left(ib\xi^2\right)\right] \exp(-iK\xi)d \xi,
\end{align}
where
\[
b=\frac{\alpha}{2} \cot\left(\alpha z\right),\qquad K=\alpha
x\csc\left(\alpha z\right),
\]
and
\begin{equation*}
f(x,z)= \sqrt{-\frac{i}{2\pi} \frac{K}{x}} \exp\left(ibx^2\right).
\end{equation*}
One can see that the integral in Eq. (\ref{eq7}) is a FT of
$\psi(x,0)\exp\left(ibx^2\right)$. Therefore, after choosing an
input $\psi(x,\,0)$, one can get an analytical evolution solution by
finding the FT of $\psi(x,0)\exp\left(ibx^2\right)$. In other words,
the propagation of a beam in a parabolic potential is equivalent to
a kind of automatic FT, that is, to the periodic change from the
beam to the FT of the beam with a parabolic chirp and back. It is
worth mentioning that Eq. (\ref{eq7}) is a fractional FT of the
initial beam
\cite{ozaktas.book.2001,mendlovic.josaa.10.1875.1993,mendlovic.ao.33.6188.1994},
the ``degree'' of which is proportional to the propagation distance.

\subsection{Hermite-Gauss beams}

Corresponding to the case with
$\psi(x,0)=\exp(-x^2/2\sigma^2)H_n(x/\sigma)$, that is, the HG beam,
the solution can be written as
\begin{align}\label{eq8}
  \psi(x,z) = f(x,z)
    \int_{-\infty}^{+\infty} \exp\left(-\beta \xi^2\right) H_n\left(\frac{\xi}{\sigma}\right) \exp(-iK\xi) d\xi,
\end{align}
where $\beta=1/(2\sigma^2) -ib$. Different from the fundamental HG
beam, the general solution of Eq. (\ref{eq8}) is nontrivial.
However, (i) when $z=m\mathcal{D}_s/2$ with $m$ an odd integer, the
solution in Eq. (\ref{eq8}) is reduced to Eq. (\ref{eq3a}), the FT
of the initial beam; (ii) given certain integer $n$, one can
calculate the corresponding analytical solution. Specifically, if
$n=0$, we have
\begin{subequations}
\begin{align}\label{eq8a}
  \psi(x,z) = & f(x,z) \sqrt{\frac{\pi}{\beta}} \exp\left(-\frac{K^2}{4\beta}\right) \notag \\
              = & f(x,z)
                \frac{\sqrt{2\pi}\sigma}{\sqrt{1-i\sigma^2\alpha\cot\left(\alpha z\right)}}
\exp\left(-\frac{\alpha^2 \sigma^2x^2\csc^2\left(\alpha z\right)}{2-i2\sigma^2\alpha\cot\left(\alpha z\right)}\right),
\end{align}
which corresponds to Fig. \ref{fig1}(a).
From Eq. (\ref{eq8a}), one can find that the beam envelope is periodic in $z$,
and the period is $\pi/\alpha$, which is in accordance with Eq. (\ref{eq4a}).

\subsection{Bessel-Gauss beams}

As concerns the BG beams, we also begin with the case of an arbitrary order $\psi(x,\,0)=\exp(-x^2/2\sigma^2)J_n(x)$.
According to Eq. (\ref{eq7}), we have
\begin{align}
  \psi (x,z)\label{eq8b}
               =  (-i)^n \frac{f(x,z)}{\beta\sqrt{\pi}}
                  \int_{-\pi/2}^{\pi/2} T_n(\sin\theta) \exp\left(-\frac{(K-\sin\theta)^2}{4\beta}\right) d\theta .
\end{align}
Comparing Eq. (\ref{eq8b}) with Eq. (\ref{eq3b}), we find that they
share the same oscillation period $\mathcal{D}_s$, hence Eq.
(\ref{eq8b}) will reduce to Eq. (\ref{eq3b}) at an odd integer
multiple of $\mathcal{D}_s/2$.

\subsection{Finite-energy Airy beams}

For the case shown in Fig. \ref{fig1}(c), one can also obtain the
corresponding analytical solution with the input $\psi(x,0)={\rm
Ai}(x)\exp(ax)$:
\begin{align}\label{eq8c}
  \psi(x,z) = & - f(x,z) \sqrt{i\frac{\pi}{b}} \exp\left(\frac{a^3}{3}\right)
                          {\rm Ai}\left(\frac{K}{2b} - \frac{1}{16b^2} + i \frac{a}{2b} \right) \times \notag \\
                &           \exp\left[
                                      \left(a+\frac{i}{4b}\right)\left(\frac{K}{2b}-\frac{1}{16b^2}+i\frac{a}{2b}\right)
                                \right]
                          \exp\left[-i\frac{K^2}{4b}
                                      -\frac{1}{3}\left(a+\frac{i}{4b}\right)^3
                                \right].
\end{align}
\end{subequations}
Taking into account the asymmetry of the finite energy Airy beam,
the period is $2\mathcal{D}_s\equiv\mathcal{D}_{\rm as}$. Note that
Eq. (\ref{eq8c}) is not valid when $z=(2m-1)\mathcal{D}_{\rm as}/4$,
because at these positions $b=0$. To obtain an analytical solution
at $z=(2m-1)\mathcal{D}_{\rm as}/4$ one must directly solve Eq.
(\ref{eq7}), which gives the result reported in Eq. (\ref{eq3c}).
Therefore, the solution for this case is a combination of Eqs.
(\ref{eq3c}) and (\ref{eq8c}).



\section{Self-Fourier beams}\label{fourier}

By now, it is apparent that the propagation of beams according to
the linear Schr\"odinger equation with parabolic potential is
intimately connected with the self-Fourier (SF) transform. It is
easy to find a beam, the FT of which is the beam itself -- the
Gaussian beam. Except Gaussians, other nontrivial functions have
been found
\cite{cincotti.jpa.25.l1191.1992,caola.jpa.24.l1143.1991,lipson.josaa.10.2088.1993,horikis.josaa.23.829.2006,
lohmann.josaa.9.2009.1992, banerjee.josaa.12.425.1995}, such as HG
functions and comb functions. In Eq. (\ref{eq8}), $b=0$ when
$z=m\mathcal{D}_s/2$, with $m$ being an odd integer, so that the
equation is just a Fourier transform of HG beam without a parabolic
chirp, and the solution is shown in Eq. (\ref{eq3a}), which is also
a HG beam. In fact, there are infinitely many SF beams, which are
constructed according to the rule $f(x)=g(x)+g(-x)+{\hat g}(x)+{\hat
g}(-x)$, where $g(x)$ is any Fourier--transformable function. This
general rule was established a while ago by Caola
\cite{caola.jpa.24.l1143.1991} and the topic of SF beams is not new
at all, as witnessed by the references cited above; still, we
believe that the method on how to prepare novel SF beams introduced
in this paper is not reported before and worthy of attention.

From Fig. \ref{fig2}, one can see that the intensity profiles in
real and $k$ spaces are the same in between the points
$z=m\mathcal{D}_s$ and $z=(2m+1)\mathcal{D}_s/2$, where $m$ is a
non-negative integer. Considering that the system is linear and
reciprocal, the places of interest should be
$z=(2m+1)\pi/(4\alpha)$. For simplicity, we choose $m=0$. When the
beam propagates to $z=\pi/4$, we have $b=\alpha/2$,
$K=\sqrt{2}\alpha x$, and
\[
f(x)= A \exp\left(i\frac{\alpha}{2}x^2\right),~
{\rm with}~
A=\sqrt{-i\frac{\alpha}{\sqrt{2}\pi}}
.\]
Therefore, Eq. (\ref{eq7}) can be rewritten as
\begin{align}\label{eq9}
  \psi(x)  =  A\exp\left(i\frac{\alpha}{2}x^2\right)
              \int^{+\infty}_{-\infty} \psi(\xi) \exp\left(i\frac{\alpha}{2}\xi^2\right)\exp\left(-i\sqrt{2}\alpha x\xi\right) d \xi .
\end{align}
We let $g(x)=\exp(i\alpha x^2/2)$, and introduce the FT operator
$\mathcal{F}[\circ](\bullet)$, in which $\circ$ and $\bullet$
represent the original function and the spatial frequency,
respectively. Equation (\ref{eq9}) can be recast as
\begin{equation}\label{eq10}
  \psi(x) = A g(x) \left\{\mathcal{F}[\psi(\xi)](K) \star \mathcal{F}[g(\xi)](K)\right\},
\end{equation}
where $\star$ represents the convolution operation,
and the spatial frequency is $K$.
Taking the FT of Eq. (\ref{eq10}) with spatial frequency $k$, one finds
\begin{align}\label{eq11}
  & \mathcal{F} [\psi(x)](k) \notag \\
  & = A \mathcal{F}[g(x)](k) \star
   \mathcal{F}[\mathcal{F}[\psi(\xi)](K) \star \mathcal{F}[g(\xi)](K)](k) \notag \\
  & = \frac{A}{8\alpha^2\pi^2} \mathcal{F}[g(x)](k) \star
 \left[ \psi\left(-\frac{k}{\sqrt{2}\alpha}\right) g\left(-\frac{k}{\sqrt{2}\alpha}\right) \right].
\end{align}
After some algebra, Eq. (\ref{eq11}) is rewritten in the form
\begin{align}\label{eq12}
  \mathcal{F}  [\psi(x)](k) =  \frac{A}{8\alpha^2\pi^2} \sqrt{\frac{2i\pi}{\alpha}}
   \left[ \exp\left(i\frac{\alpha}{2}\xi^2\right) \psi(\xi)\right] \star \exp\left(- i \alpha \xi^2 \right),
\end{align}
in which we let $k=-\sqrt{2}\xi\alpha$.
The convolution in Eq. (\ref{eq12}) can be written as
\begin{align*}
  \frac{1}{2\pi}  \exp\left(-i\alpha x^2\right)
   \int_{-\infty}^{+\infty} \psi(\xi) \exp\left(-i\frac{\alpha}{2}\xi^2\right) \exp\left(i2\alpha x\xi\right)d\xi.
\end{align*}
If one lets $x=ix'/\sqrt{2}$ and $\xi=i\xi'$, the convolution is finally put in the form
\begin{align}\label{eq13}
\frac{i}{2\pi} \exp\left(i\frac{\alpha}{2}x'^2\right)
 \int^{+\infty}_{-\infty} \psi(\xi') \exp\left(i\frac{\alpha}{2}\xi'^2\right)\exp\left(-i\sqrt{2}\alpha x'\xi'\right) d \xi' ,
\end{align}
which is just the integral in Eq. (\ref{eq9}). Therefore, the beam
envelopes in real and inverse spaces have the same profile (up to a
transverse scaling), i.e. the beams represented by Eq. (\ref{eq9})
are the SF beams. In light of the fact that the initial beam can be
arbitrary, the number of SF beams is infinite and can be easily
made. \textit{Generally, for an arbitrary $\psi(x)$ propagating to
$\pi/(4\alpha)$ in  a parabolic potential, a self-Fourier beam will
be obtained.} Note that if the initial beam is parity-asymmetric,
the corresponding FT still has the same profile, but with an
intermediate inversion. In general, the corresponding FT pair is
\begin{equation}\label{eq14}
  \mathcal{F}[\psi(x)](k) = \sqrt{\frac{2\pi}{\alpha}} \psi\left(-\frac{k}{\alpha}\right).
\end{equation}
We would like to emphasize that this method is completely different
from the previously reported methods. Its discovery is enabled by
the useful properties of beam propagation in a parabolic potential.

\begin{figure}[htbp]
\centering
  \includegraphics[width=0.5\columnwidth]{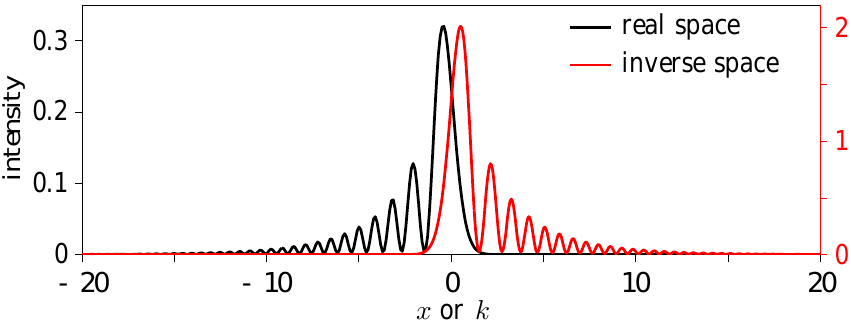}
  \caption{(Color online) Comparison of intensities of an AB at $z=\pi/4$ in real space and inverse space,
  corresponding to Fig. \ref{fig2}(a).
  Intensities in real and frequency spaces refer to the left and right $y$ scales, respectively.
  }
  \label{fig4}
\end{figure}

As an example, we discuss the finite energy Airy beam $\psi(x,z=0)={\rm Ai}(x)\exp(ax)$ with $a=0.1$
as an initial beam in a parabolic potential.
Plugging this beam in Eq. (\ref{eq7}), one obtains the corresponding SF beam at $z=\pi/(4\alpha)$, written as
\begin{align}\label{eq10}
\psi(x) =  & - \sqrt[4]{2} \, {\rm Ai}\left(\sqrt{2}x - \frac{1}{4\alpha^2} + i \frac{a}{\alpha} \right)
           \exp\left[a
                                      \left(\sqrt{2}x-\frac{1}{2\alpha^2}\right)
                          \right] 
            \exp\left[
                                      -\frac{i}{2}\left(2\alpha x^2-\frac{\sqrt{2}}{\alpha}x+\frac{a^2}{\alpha} +\frac{1}{6\alpha^3}\right)
                         \right].
\end{align}
In Fig. \ref{fig4}, the intensity of the SF beam is shown by the
black curve, and the corresponding intensity in Fourier space is
shown by the red curve. One can see that the beam profiles are the
same except for an inversion, which is in accordance with the
theoretical result presented in Eq. (\ref{eq14}).

\begin{figure}[htbp]
\centering
  \includegraphics[width=0.5\columnwidth]{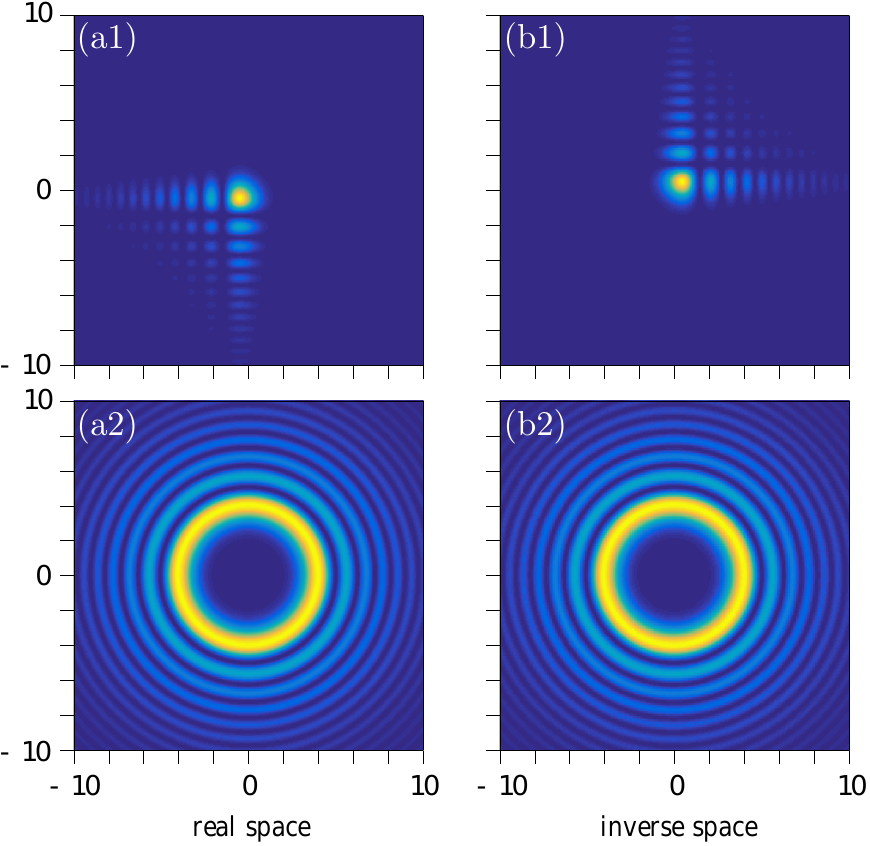}
  \caption{(Color online) (a1) Intensity of a two-dimensional finite-energy Airy beam ${\rm Ai}(x){\rm Ai}(y)\exp[a(x+y)$ at $z=\pi/4$ in real space.
  (b1) The corresponding intensity in inverse space.
  (a2) and (b2) Same as (a1) and (b1), but for a circular finite-energy Airy beam $\psi(x,y,z=0)={\rm Ai}(r_0 - r)\exp[a(r_0 - r)]$.
  The parameters are $\alpha=1$, $r_0=5$, and $a=0.1$.
  }
  \label{fig5}
\end{figure}

Naturally, the harmonic oscillator model in Eq. (\ref{eq1}) is
easily extended to two, three or even four dimensions. The hydrogen
atom in three dimensions can be represented as a four-dimensional
harmonic oscillator \cite{cornish.jpa.17.323.1984}. Specifically, in two
dimensions the problem can be reduced by the separation of variables
to two one-dimensional cases \cite{zhang.oe.23.10467.2015}. In Fig.
{\ref{fig5}}(a1) (real space) and \ref{fig5}(b1) (inverse space), we
display the results for a two-dimensional finite-energy Airy beam
\cite{siviloglou.ol.32.979.2007}
\[
\psi(x,y,z=0)={\rm Ai}(x){\rm Ai}(y)\exp[a(x+y)].
\]
As expected, the beam at $z=\pi/4$ is a SF beam, and the
corresponding analytical solution can be obtained immediately from
Eq. (\ref{eq10}) -- by making a product of $x$ and $y$ components.

In addition to the inputs with separated variables, one can also
treat two-dimensional inputs with non-separable variables, for
example, an outward circular finite-energy Airy beam
\cite{zhang.ol.36.2883.2011,chremmos.ol.36.3675.2011}. Such an
initial beam can be written in polar coordinates as
\[
\psi(x,y,z=0)={\rm Ai}(r_0 - r)\exp[a(r_0 - r)],
\]
where $r_0$ determines the location of the main ring. As shown in
Figs. \ref{fig5}(a2) and \ref{fig5}(b2), the beam at $z=\pi/4$ is
still a SF beam. However, the corresponding explicit analytical
solution cannot be obtained, which is true for most of the initial
beams. Still, the method is valid for all kinds of initial beams,
regardless of whether the results can be expressed explicitly or
not.

\section{Conclusion and outlook}\label{conclusion}

In conclusion, we have theoretically and numerically investigated
the beam propagation in a linear medium with an external parabolic
potential. According to this model, the beam will undergo profile
oscillation during propagation, which will suppress diffraction. For
an asymmetric beam, such as the finite energy AB, there is a
periodic inversion during propagation. The oscillation period for
parity-symmetric beams is half that of parity-asymmetric beams.

We also find that an arbitrary beam realizes automatic FT during
propagation. At half the period for parity-symmetric beams and a
quarter of the period for parity-asymmetric beams, the FT of the
initial beams is obtained repeatedly; at the places in between, the
beam is the FT of the initial beam with a parabolic chirp.

Last but not least, based on the theoretical model, we discover a
class of SF beams. This method may be easily implemented in
experiment -- one can use a GRIN medium or a Lohmann optical system
\cite{goodman.book, ozaktas.book.2001}, which are extensively
investigated and well understood. The properties discussed not only
exhibit the importance of parabolic potential in linear optics, but
broaden the understanding of recurrence in the paraxial beam
propagation, and also exemplify scientific significance of the
parabolic potential propagation for signal processing, imaging,
microparticle manipulation, information storage, and other
applications.

In the end, we mention some of the useful extensions of the theory
presented here. In addition to GRIN media, to which the theory
applies directly, Bose-Einstein condensates with harmonic traps \cite{mihalache.rjp.59.295.2014,bagnato.rrp.28.251.2015} are
viable candidates, when the nonlinearity is weak
\cite{skupin.pre.70.016614.2004, zhang.oe.18.27846.2010}. Talbot
effect in one and two dimensions can also be presented as a SF
phenomenon -- in fact, as a fractional FT
\cite{lohmann.josaa.9.2009.1992, mendlovic.josaa.10.1875.1993}. A
big challenge would be to extend these ideas to nonlinear domain --
which in principle is difficult, since FT by itself is a linear
operation. Nonetheless, we have recently demonstrated
\cite{zhang.pre.89.032902.2014, zhang.pre.91.032916.2015} that the
nonlinear Talbot effect can be presented in terms of primary and
secondary recurrences, which might be interpreted as SF phenomena.
Quite surprisingly, it is also proven that the hyperbolic secant
function -- one of the workhorses in soliton theory -- is a SF
function \cite{banerjee.josaa.12.425.1995}.

\section*{Acknowledgments}
This work was supported by the 973 Program (2012CB921804), KSTIT of
Shaanxi province (2014KCT-10), NSFC (61308015, 11474228), NSFC of
Shaanxi province (2014JQ8341), CPSF (2014T70923, 2012M521773), and
the NPRP 6-021-1-005 project of the Qatar National Research Fund (a
member of the Qatar Foudation).

%

\end{document}